# Stimulated Raman Scattering in Nonlinear Silicon Nanophotonic Waveguides: Theory and Applications in Photonic Integrated Circuits

Abdurrahman Javid Shaikh[1*] and Othman Sidek[2]

[1,2] School of Electrical & Electronic Engineering, Universiti Sains Malaysia, Nibong Tebal 14300, Malaysia (ajs11_eee138@student.usm.my) * Corresponding author

**Abstract**: Photonics caught world attention since channel capacity limit of metallic interconnects approached due to research and design in high speed digital processors. Use of dielectrics, instead, suitable for light propagation was more attractive due to its extremely wide bandwidth. Many of the devices, both active and passive, have been demonstrated using these insulating materials. Due to its excellent optical characteristics, established fabrication history, and cheaper throughput, silicon found its place in photonics arena. However, due to its indirect band structure, efficient light sources are not possible using silicon as the base material. Nevertheless, techniques such as stimulated Raman scattering and third-harmonic generation have made it possible to avoid this natural hurdle in the path of silicon as a light source. This paper reviews basic theory of stimulated Raman scattering, its applications in the context of silicon based photonic integrated circuits and describes ways to improve this nonlinear effect. This paper also covers few of the most important demonstrations of stimulated Raman scattering published in literature from the last decade.

## I. INTRODUCTION

There are two parameters which incessantly improved – or were required to be improved – as the electronic communication market rose exponentially. This rise in demand is naturally distilled from the continually changing life style of our global society as a whole. Specifically, ubiquitous footprint of smart devices, surveillance infrastructure, and upcoming internet of things (IoT) necessitated this improvement. The two parameters are: (i) the device size which continuously decreased in pursuance of the Moore's law, and (ii) communication bandwidth whether the communication be wireless or otherwise. Moore's law defined a trend in microelectronic industry that number of transistors that can be fabricated on a chip will double after every two years. This law is not only a technological trend but it has economical repercussions as well.

Many materials have been playing their role in the electronic industry but nothing compares silicon in terms of performance and economics together. The half-century long understanding of silicon material and its processing has manifested itself in a mature VLSI fabrication infrastructure. Currently this infrastructure is capable of flushing markets with multi-core microprocessors capable of handling data at multiples of gigahertz. The dramatically improved speed of these processors is actually limited by metallic interconnects due to their relatively low bandwidth by virtue of finite resistance, capacitance and skin effect [1]. Consequently, to optimize overall performance, channel capacity of the interconnects has to be increased. What if these metallic conductors can be replaced by insulators and instead of electronic signals which employ electron transport, optical signals which employ photon travel at the speed of light are used? A lot of work has been done to investigate this question, which falls under the umbrella of all-optical systems and photonic integrated circuits (PICs) which has shown promise of virtually infinite bandwidth and device miniaturization up to nano-scale [2].

The principal components of photonic systems are optical sources with ultrafast modulation, appropriate transmission media and optical detectors. Appropriate transmission media may include amplification units, either inherent amplification by the media itself or separately done. Silicon material has been successfully demonstrated for high speed modulation and switching [3-8], sensing and detection [9-10], pulse shapers [11-12], and wavelength converters [13-15]. Optical amplification and lasing are the two functionalities faced many difficulties mainly due to indirect band-gap of silicon. Material properties of silicon such as high optical damage threshold of 1-4 GW/cm$^2$ [2] as well as five order of magnitude larger Raman gain coefficient as compared to glass fibers encouraged realization of optical amplification [16-17] and lasing [18-20] using stimulated Raman scattering (SRS). Following, we briefly discuss Raman scattering before taking up SRS and its applications in the context of photonic integrated circuits in the sections to follow.

## II. RAMAN SCATTERING

Raman scattering, like Kerr nonlinearity, can only be observed substantially with high intensity optical fields interacting with suitable material (Raman medium). Spontaneous Raman Scattering, observed in 1928, is a nonlinear inelastic optical process. In this process, a photon from a high intensity optical source (a "pump" at frequency $\omega_P$) excites bound carriers of a crystal to an unstable state which immediately settle back to a more stable state by emitting an optical photon. The emitted photon may have wavelength longer ($\omega_S$, Stokes wavelength) or shorter ($\omega_{AS}$, anti-Stokes wavelength) than the wavelength of pump photon depending upon the Raman spectrum of the crystal material. If the Stokes photon is emitted, the photon energy ($\hbar\omega_S$) is lower than pump energy ($\hbar\omega_P$) and the medium is left excited as shown in Fig. 1(a) (red represents higher energy and blue shows lower). If the medium is already excited then reverse scattering can take place whereby anti-Stokes photon is emitted with higher photon energy ($\hbar\omega_{AS}$) than pump energy ($\hbar\omega_P$). In this case, the medium is now less excited than before scattering occurred, as shown in Fig. 1(b) (red represents higher energy and blue shows lower). These frequencies are given by;

$\omega_S = \omega_P - \omega_R$, (Stokes Line), (1)
$\omega_{AS} = \omega_P + \omega_R$, (Anti-Stokes Line), (2)

where, $\omega_R$ is the Raman shift which depends upon scattering material. Typically, Stokes lines dominate the emission because ground level states are much more in number than excited states, in thermal equilibrium. Moreover, Stokes line's power is proportional to the pump intensity therefore can be controlled up to certain limits.

## III. STIMULATED RAMAN SCATTERING

In Stimulated Raman Scattering, two optical beams, namely, pump (at high intensity) and probe (tuned at Stokes shifted frequency) are used. The pump creates population inversion within the medium while the low-intensity probe photon passing by an excited atom/molecule stimulates it to emit another Stokes photon with exactly the same energy, phase and direction. This results in highly collimated and coherent beam with extremely narrow linewidth. In essence, energy from high intensity pump is transferred to the emitted beam at Stokes wavelength. The pump frequency is typically kept well below the band gap frequency so that generation of free carriers and thereby free carrier absorption (FCA) could be avoided. However, pumping intensity required to observe Raman process (known as pump threshold) is lower when high energy pump is used, and since high optical intensity can damage Raman medium, there is a trade-off that needs to be made. Pump properties are extremely important in Raman setup as Raman linewidth is proportional to pump spectrum [21]. The Stokes shift frequency ($\omega_R/2\pi$) of silicon is 15.6THz with Full-Width Half-Maximum (FWHM) of emitted spectrum is only 105GHz centered at 15.7THz below pump frequency [22-24]. Hence for amplification using Raman effect, only a narrow bandwidth is available. As net gain depends on the signal wavelength relative to the spectrum center, SRS can be used to selectively amplify optical signals at various wavelengths.

Use of silicon as the Raman medium to observe spontaneous Stokes and Anti-Stokes emissions was first reported in 2002 [25] with Raman scattering efficiency as low as $(4.1 \pm 2.5) \times 10^{-7}$ cm$^{-1}$ Sr$^{-1}$ in a 2.4cm long waveguide (WG). Recently, Takahashi et al. [26], reported their encouraging experiment which used high quality photonic crystal silicon nanowire to enhance spontaneous Raman scattering and showed that it is proportional to the quality of the nanowire. SRS in silicon WG, however, was first reported by the same group (i.e. of [25]) in 2003 [16] which reported Stokes signal amplification at 1542.3nm using 1427nm pump to excite 1.8cm long WG having an effective area of 5.4µm$^2$. In the same year, wavelength conversion was also reported via Coherent Anti-Stokes Raman Scattering (CARS) [27]. Soon after reports of amplification and wavelength conversion, a pulsed silicon Raman laser was reported in 2004 [18] which was followed by reports of continuous-wave operation in 2005 [20].

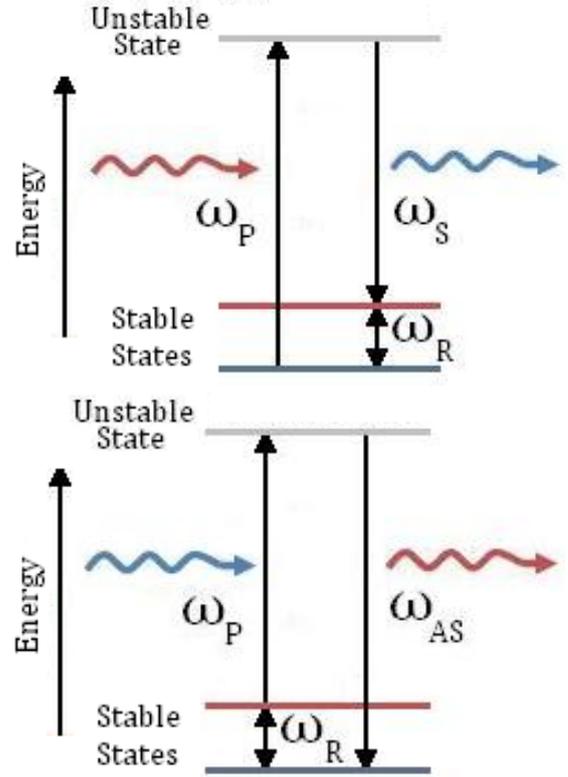

Figure 1. (a) Raman scattering at Stokes frequency; (b) Raman scattering at anti-Stokes frequency.

A more recent report presented Raman amplification of data signal at 40 Gb/s in a low-loss silicon WG [28]. The report noted field amplification in a signal at wavelength of 1544.9nm inside an S-shaped, 4.6cm long, rib waveguide with bend radii of 400µm when it was pumped with 1430nm high intensity beam. A maximum on/off gain of 2.3 dB was observed with a coupled pump power of 230mW. At higher coupled pump power the amplification was expected to increase but not appreciably because of saturation in the curve (gain vs couple power). Both co-propagating and counter-propagating pump and probe signals were analyzed. Though, on-off gain (OOG) and signal-to-noise ratio (SNR) remained the same without pump in both cases (6.6dB), in co-propagating (counter-propagating) configuration, however, when pump was applied to amplify the signal a reduced SNR of 4.9dB (5.5dB) was observed corresponding to a noise figure (NF) of 1.3dB (0.8dB). These observations suggest improved SRS observation in counter-propagating case (due to lower NF and no jittering) as counter propagating beams reduced noise share from Four Wave Mixing (FWM) and Cross Phase Modulation (XPM). Raman gain, on the other hand, is independent of the relative direction of propagation [28]. Rong, Xu et al., in 2007 [29], made a big leap in Raman based silicon laser by reporting a very low-threshold of only 20mW with continuous-wave (cw) operation at Stoke's wavelength of 1686nm. The fabricated device consisted of a racetrack p-i-n rib waveguide (see Fig. 2) of 3cm in length with bend radius of 400μm. Pump and probe signal were coupled to the waveguide using directional coupler, critically tuned for pump wavelength of 1550nm while offering low coupling ratio for probe signal. The pump and probe fields were coupled into the WG through a lensed fiber. Reverse biasing the p-i-n structure with 25V, a maximum output power of 50mW was observed with pump power of ~290mW. This demonstration shows a ten-fold improvement over the Raman lasers demonstrated earlier [20, 29]. In both the reports [28-29], a p-i-n (p type-intrinsic-n type) waveguide was used to actively remove free-carriers out of the intrinsic modal region which effectively reduced carrier lifetime thereby increased pump efficiency and WG quality factor. Fig. 3(a) shows an SEM image of p-i-n diode waveguide used for Raman amplification and lasing experiments [21], while, Fig. 3(b) shows the SEM image of cross-section of directional coupler fabricated for experiments in [29]. Achieving optical gain and lasing using SRS made it possible to avoid demerits of indirect band structure of silicon.

Typically very long WGs are required to observe nonlinear interactions resulting in Raman gain, which required kilometers of length for silica fibers. Silicon, on the other hand, has around four-order of magnitude greater SRS gain coefficient as well as tight single mode confinement in relatively very small cross-section area due to large index contrast in silicon-on-insulator (SOI) platform, as compared to single mode silica fibers, made realization of SRS amplifiers and lasers possible in cm long WGs. Net optical gain in low-loss SOI WG by SRS has already been reported, in case of pulsed pump operation (2dB [30]) and continuous-wave pump operation (>3dB [31]), in a 4.8cm long WG. However, these lengths are still not ideal for on-chip communications where CMOS compatibility requires WG length in micrometers. Earlier reported result of cw spontaneous Raman gain in a 4.2mm long SOI nanowire WG with cross-section of 0.098μm2 gave encouraging value of 0.7dB [32]. Nonetheless, results of [30-31] are suitable for Raman based discrete amplifiers where pumping the cladding of WGs, instead of core pumping, may offer high optical gain by avoiding Two Photon Absorption (TPA)-induce-FCA loss in the core [33].

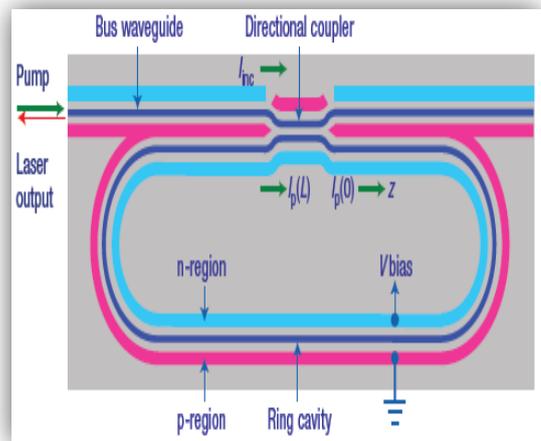

Figure 2. Schematic layout of the silicon racetrack p-i-n rib waveguide structure with heavily doped p and n regions. Reprinted by permission from Macmillan Publishers Ltd: Nature Photonics [29], copyright 2007

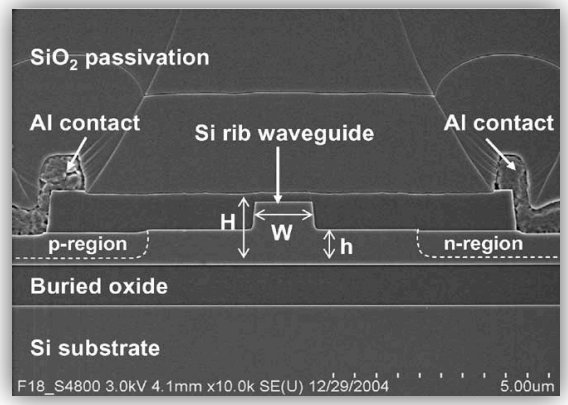

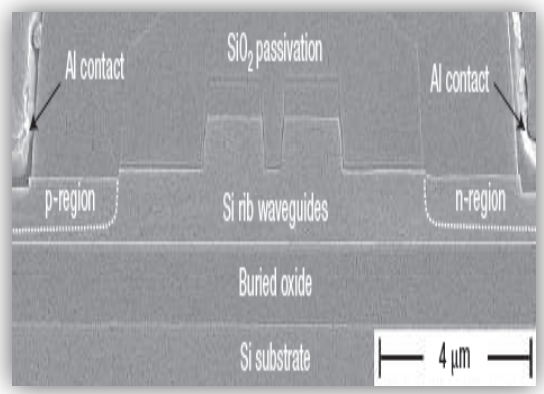

Figure 3. Scanning Electron Microscope (SEM) image of, (a)*: r-i-b waveguide; (b)**: directional coupler region of the racetrack waveguide.
* Copyright © 2006 IEEE. All rights reserved. Reprinted, with permission, from [21].
** Reprinted by permission from Macmillan Publishers Ltd: Nature Photonics [29], copyright 2007.

Like any nonlinearity, SRS in silicon can be enhanced (i) by making the optical field travel slow inside the WGs (slow light) [34], and (ii) by increasing optical field confinement [35]. The former increases probability of photon interaction with WG crystals while later enhances the electric field inside the WG. Increased electric field inside a WG means increased energy density which, in turn, means occurrence of nonlinear phenomena at much reduced input power (i.e., low-threshold). This high confinement is one of the design parameters for maximum nonlinear optical interaction in WGs [22] – thus achieving high quality medium. This high quality is required to observe Raman effect in optical amplifiers [16] and lasers [19-20]. Whether it is high optical field confinement or achieving slow light, large index contrast and large refractive index of WG material are keys to achieve them. The high optical field confinement within SOI WG is understood due to large step between indices of silicon core and insulator ($SiO_2$) cladding ($\Delta n \sim 2.1$). Besides, tight optical field confinement allows fabrication of WGs with ultra-compact bending of radius in micron range [36] hence making the device CMOS-compatible. A tremendously improved solution to enhance slow light effect is to use photonic crystal silicon waveguide which also offers tight optical field confinement [34].

## V. CONCLUSION

In this paper, we presented basic theory of Raman scattering. Nonlinear behavior of silicon can be enhanced by achieving high figure of merit which is achieved by high optical field confinement and spatial compression of optical field in photonic crystal waveguides. Nevertheless, silicon is not a natural optical source for its indirect band structure, therefore optical gain and lasing seems impossible using conventional methods. Simulated Raman scattering has been demonstrated to show these functionalities while circumventing restriction posed by indirect band of silicon. Though optical pumping has been successfully demonstrated using Raman scattering, efficiency and power dissipation are not satisfactory [1]. Besides, optically pumped sources are not the ideal ones for all-optical PICs as they require external lenses to be fabricated to couple pump light into WGs, thereby, limiting density of integration. Nevertheless, SRS based amplification and lasing is one of the most attractive solutions for discrete or off-the-chip applications. Hence optical sources based on silicon are highly expected to be built based on silicon which is a requirement of all-optical silicon based photonic circuits.

## REFERENCES

[1] D. J. Paul, "Silicon photonics: a bright future?," *Electronics Letters,* vol. 45, pp. 582-584, 2009.
[2] B. Jalali and S. Fathpour, "Silicon Photonics," *J. Lightwave Technol.,* vol. 24, pp. 4600-4615, 2006.
[3] V. R. Almeida, *et al.*, "All-optical switching on a silicon chip," *Opt. Lett.,* vol. 29, pp. 2867-2869, 2004.
[4] F. Y. Gardes, *et al.*, "40 Gb/s silicon photonics modulator for TE and TM polarisations," *Opt. Express,* vol. 19, pp. 11804-11814, 2011.
[5] W. M. Green, *et al.*, "Ultra-compact, low RF power, 10 Gb/s siliconMach-Zehnder modulator," *Opt. Express,* vol. 15, pp. 17106-17113, 2007.
[6] L. Liao, *et al.*, "High speed silicon Mach-Zehnder modulator," *Opt. Express,* vol. 13, pp. 3129-3135, 2005.
[7] A. Liu, *et al.*, "A high-speed silicon optical modulator based on a metal-oxide-semiconductor capacitor," *Nature,* vol. 427, pp. 615-618, 2004.
[8] Q. Xu, *et al.*, "Micrometre-scale silicon electro-optic modulator," *Nature,* vol. 435, pp. 325-327, 2005.
[9] F. Dell'Olio and V. M. Passaro, "Optical sensing by optimized silicon slot waveguides," *Opt. Express,* vol. 15, pp. 4977-4993, 2007.
[10] J. T. Robinson, *et al.*, "On-chip gas detection in silicon optical microcavities," *Opt. Express,* vol. 16, pp. 4296-4301, 2008.
[11] R. Salem, *et al.*, "All-optical regeneration on a silicon chip," *Opt. Express,* vol. 15, pp. 7802-7809, 2007.
[12] E.-K. Tien, *et al.*, "Pulse compression and modelocking by using TPA in silicon waveguides," *Opt. Express,* vol. 15, pp. 6500-6506, 2007.
[13] R. Espinola, *et al.*, "C-band wavelength conversion in silicon photonic wire waveguides," *Opt. Express,* vol. 13, pp. 4341-4349, 2005.
[14] M. A. Foster, *et al.*, "Broad-band continuous-wave parametric wavelength conversion insilicon nanowaveguides," *Opt. Express,* vol. 15, pp. 12949-12958, 2007.
[15] Y.-H. Kuo, *et al.*, "Demonstration of wavelength conversion at 40 Gb/s data rate in silicon waveguides," *Opt. Express,* vol. 14, pp. 11721-11726, 2006.
[16] R. Claps, *et al.*, "Observation of stimulated Raman amplification in silicon waveguides," *Opt. Express,* vol. 11, pp. 1731-1739, 2003.
[17] M. A. Foster, *et al.*, "Broad-band optical parametric gain on a silicon photonic chip," *Nature,* vol. 441, pp. 960-963, 2006.
[18] O. Boyraz and B. Jalali, "Demonstration of a silicon Raman laser," *Opt. Express,* vol. 12, pp. 5269-5273, 2004.
[19] A. L. H. Rong, R. Jones, 0. Cohen, D. Hak, R. Nicolaescu, A. Fang, and M. Paniccia,, "An all-silicon Raman laser," *Nature,* vol. 7023, pp. 292-294, 2005.
[20] H. Rong, *et al.*, "A continuous-wave Raman silicon laser," *Nature,* vol. 433, pp. 725-728, 2005.
[21] L. Ansheng, *et al.*, "Optical amplification and lasing by stimulated Raman scattering in silicon waveguides," *Lightwave Technology, Journal of,*

vol. 24, pp. 1440-1455, 2006.
[22] J. Leuthold, *et al.*, "Nonlinear silicon photonics," *Nat Photon,* vol. 4, pp. 535-544, 2010.
[23] H. Rong, *et al.*, "Raman gain and nonlinear optical absorption measurements in a low-loss silicon waveguide," vol. 85, pp. 2196-2198, 2004.
[24] L. Tak-Keung and T. Hon-Ki, "Nonlinear absorption and Raman scattering in silicon-on-insulator optical waveguides," *Selected Topics in Quantum Electronics, IEEE Journal of,* vol. 10, pp. 1149-1153, 2004.
[25] R. Claps, *et al.*, "Observation of Raman emission in silicon waveguides at 1.54 μm," *Opt. Express,* vol. 10, pp. 1305-1313, 2002.
[26] Y. Takahashi, *et al.*, "First observation of Raman scattering emission from silicon high-g photonic crystal nanocavities," in *Lasers and Electro-Optics (CLEO), 2011 Conference on*, 2011, pp. 1-2.
[27] R. Claps, *et al.*, "Anti-Stokes Raman conversion in silicon waveguides," *Opt. Express,* vol. 11, pp. 2862-2872, 2003.
[28] V. Sih, *et al.*, "Raman amplification of 40 Gb/s data in low-loss silicon waveguides," *Opt. Express,* vol. 15, pp. 357-362, 2007.
[29] H. Rong, *et al.*, "Low-threshold continuous-wave Raman silicon laser," *Nat Photon,* vol. 1, pp. 232-237, 2007.
[30] A. Liu, *et al.*, "Net optical gain in a low loss silicon-on-insulator waveguide by stimulated Raman scattering," *Opt. Express,* vol. 12, pp. 4261-4268, 2004.
[31] R. Jones, *et al.*, "Net continuous wave optical gain in a low loss silicon-on-insulator waveguide by stimulated Raman scattering," *Opt. Express,* vol. 13, pp. 519-525, 2005.
[32] R. Espinola, *et al.*, "Raman amplification in ultrasmall silicon-on-insulator wire waveguides," *Opt. Express,* vol. 12, pp. 3713-3718, 2004.
[33] M. Krause, *et al.*, "Efficient Raman Amplification in Cladding-Pumped Silicon Waveguides," in *Group IV Photonics, 2006. 3rd IEEE International Conference on*, 2006, pp. 61-63.
[34] T. Baba, "Slow light in photonic crystals," *Nat Photon,* vol. 2, pp. 465-473, 2008.
[35] C. Monat, *et al.*, "Slow Light Enhanced Nonlinear Optics in Silicon Photonic Crystal Waveguides," *Selected Topics in Quantum Electronics, IEEE Journal of,* vol. 16, pp. 344-356, 2010.
[36] Y. Vlasov and S. McNab, "Losses in single-mode silicon-on-insulator strip waveguides and bends," *Opt. Express,* vol. 12, pp. 1622-1631, 2004.